\documentclass[twocolumn,twoside,slac_two]{revtex4}
\usepackage{graphicx}
\usepackage{fancyhdr}
\begin{document}
\title{Kaons: Review and Outlook}

%

\author{A. Ceccucci}
\affiliation{CERN, 1211 Geneva 23, Switzerland}

\begin{abstract}
This article presents a review of recent results and an outlook of kaon physics. After enjoying a renaissance, the discipline is now becoming an endangered species. Action will be needed to keep kaon physics at the heart of future FPCP meetings.  
\end{abstract}

\maketitle

\thispagestyle{fancy}


\section{Introduction}

The experiments built during the past decade to measure direct CP violation in the two pion decays of the neutral kaons ($\epsilon^\prime / \epsilon$) have renewed the interest in kaon physics. In addition to the measurement of $\epsilon^\prime / \epsilon$~\cite{Batley:2002gn,Alavi-Harati:2002ye}, these experiments have re-written the particle data tables of the neutral kaons for what concerns rare and not-so-rare kaon decays, lifetimes and form factors.
Sometimes the new measurements disagreed with the old ones, some of which have been published more than thirty years ago. For instance, in the case of semileptonic decays, the new round of measurements changed significantly the experimental input to determine the CKM parameter $V_{us}$. Progress has not been limited to neutral kaons: a decisive search for direct CP violation in the three pion decays of charged kaons was performed at CERN by the NA48/2 experiment. We will briefly describe these results, including some very interesting and unexpected by-products. Last but not least, we will review the present and future of rare kaon decays. Unfortunately, during the past couple of years, all the US-based kaon projects have been stopped. Nonetheless, new initiatives at the CERN-SPS and at the new J-PARC facility in Japan promise to exploit the remarkable physics potential offered by the study of rare kaon decays. The paper is organized as follows: in Section~\ref{semileptonic}, recent results on semileptonic kaon decays are presented. Results from the the study of charged kaons by the CERN experiment NA48/2 are described in Sections~\ref{na48/2},~\ref{cusp} and ~\ref{susy}. In Section~\ref{ckmmat} we put rare kaon decays in the broader contest of flavor physics. Progress toward the study of the very rare decay $K^0_{L} \to \pi^0 \nu \bar{\nu}$ is given in Section~\ref{E391a}. Finally, in Section~\ref{KPLUS} we review the prospects to make a definite measurement of $K^+ \to \pi^+ \nu \bar{\nu}$ at the CERN SPS. 

\section{Semileptonic Kaon Decays and $\mathbf{|V_{us}|}$}
\label{semileptonic}

The cleanest determination of the $|V_{us}| f_+(0)$ product is provided by the study of $K^0_{L} \to \pi^\pm e^\mp \nu$ decays for the following two reasons:
 
 \begin{enumerate}
 
 \item 
 Given that the electron mass is very small and can be neglected, only one form factor enters the calculation of the the decay amplitude.
 \item
  There are no isospin-breaking uncertainties coming from the $\pi^0 -\eta$ mixing  which complicate the calculation of the form factor in charged kaon decays.
 \end{enumerate}
 The master formula is:
\begin{equation}
\begin{array}{rcl} 
|V_{us}|\times f_+(0)= & & \nonumber \\ 
 &\sqrt{\frac{BR(K^0_{L} \to \pi e \nu)}{\tau_{L}}\cdot\frac{192 \pi^3}{G_F^2 M_K^5 S_{EW}(1+\delta_K)I^e_K}},  &  \nonumber\\
\end{array}
\end{equation}
where $S_{EW}= 1.023$ is the short distance electroweak correction~\cite{Sirlin:1981ie}, $\delta_K$ is the long-distance radiative correction and $G_F$ is the Fermi constant. In addition to the semileptonic branching ratio and $K^0_L$ lifetime ($\tau_{L}$), the form factor dependence is also needed as experimental input because it enters in the phase space integral ($I^e_K$) calculation.

 KTeV, NA48 and KLOE have provided measurements of the semileptonic branching ratio of the long-lived neutral kaon~\cite{Alexopoulos:2004sw,Lai:2004bt,Ambrosino:2005ec} and the results are reported in Table~\ref{Ke3}. The new values are all significantly higher than previously accepted.
 \begin{table}[h]
\begin{center}
\caption{Recent Measurements of ${\cal BR}(K^0_{L} \pi e \nu (\gamma))$.}
\begin{tabular}{|l|c|}
\hline \textbf{Experiment} & \textbf{BR (\%)} \\
\hline  
PDG04 &  $38.81 \pm 0.27$ \\
\hline
KTeV & $40.67 \pm 0.11$\\
\hline
NA48 &  $40.10 \pm 0.45$ \\
\hline
KLOE & $40.07 \pm 0.15$ \\
\hline
\end{tabular}
\label{Ke3}
\end{center}
\end{table}
What makes the KLOE results particularly interesting is the unique feature of the $\phi$ factory. The DA$\Phi$NE $e^+ e^-$ collider operates at the formation energy of the $\phi$ meson which subsequently decays into a pair of kaons. The quantum numbers of the $\phi$ meson guarantee that the pair of neutral kaons are self-tagged. Thus, the reconstruction of a $K^0_{S}$ meson on one side of the detector determines the presence of a $K^0_{L}$ on the other side and vice versa. This fact opens the possibility to study absolute branching ratios without problems related to the normalization channel. Another advantage is that the $\phi$ is almost at rest in the laboratory frame and therefore the momentum of the kaons is almost monochromatic. In turn this simplifies the reconstruction of the semileptonic decays since there are no ambiguities due to the unmeasured momentum of the escaping neutrino. 

Several measurements of the $K_{e3}$ form factor dependence has been recently published~\cite{Alexopoulos:2004sy,Lai:2004kb}. The dependence on the momentum transfered to the $e-\nu $ system $t=(P_K -P_\pi)^2$ has been described by a linear, quadratic or pole-model dependence. The quadratic dependence is expressed according to the relation:
\begin{equation}
 f_+(t) =f_+(0) \times \left( 1 + \lambda^\prime t/m_{\pi^+} + 1/2 \lambda^\prime{^\prime} (t/m_{\pi^+})^2\right)   
\end{equation} 
The parameters extracted from the quadratic form are strongly anti-correlated and there is no agreement between the quadratic terms measured by KTeV and NA48. 
Very recently, the results from KLOE~\cite{Ambrosino:2006gn} have become available:
\begin{equation}
\begin{array}{c} 
 \lambda^\prime = ( 25.5 \pm 1.5 \pm 1.0) \times 10^{-3}  \\
  \lambda^\prime{^\prime} = ( 1.4 \pm 0.7 \pm 0.4) \times 10^{-3} \\ 
  \rho(\lambda^\prime_+,\lambda^\prime{^\prime}_+)= -0.95.  
  \end{array}
\end{equation} 
These results agree with the determinations of NA48 and slightly disagree with the determination of KTeV. The alternative parametrization made using a pole model fit to the vector gives: $M_V = 870(7)$ MeV. 

A third advantage is given by the small Q value of the $\phi \to K^0_{S}K^0_L$ decay. The $K^0_L$ moves at small momentum $P_K \sim 110$ MeV/{\it c} and has a short mean free path before decaying ($\sim$ 340 cm). Thus, a large fraction of the decays are collected from the fiducial volume of the detector, opening the possibility to make a good measurement of the $K^0_{L}$ lifetime ($\tau_{L}$), an important input to the computation of $|V_{us}|$. It is worth to mention that the previous precise measurement of the $K^0_{L}$ lifetime was performed~\footnote{With D.~Bryman, our session chairman, as co-author!} more than 30 years ago~\cite{Vosburgh:1973vh}: $\tau_{L} = 51.54 \pm 0.44$ ns. The measurement of the $K^0_{L}$ lifetime can be performed by KLOE into two independent ways: either directly, reconstructing the proper time distribution of the kaons using $ K^0_{L} \to 3\pi^0$ decays~\cite{Ambrosino:2005vx}, or indirectly by requiring the sum of the branching ratios of the seven largest decay modes to be one~\cite{Ambrosino:2005ec}. The two determinations of the life-time are independent. Averaging the two measurements KLOE quotes: 
\begin{equation}
\tau_{L} = 50.84 \pm 0.23~{\rm ns},
\end{equation} 
which improves by a factor of two over the previous best result.    
Given the new precise experimental input, to answer the question whether or not the sum of the square of the elements belonging to the first row of the CKM matrix satisfies unitary  relies on which theoretical input is used for $f_+(0)$. A recent review prepared by Blucher and Marciano~\cite{pdgupdate} for the Review of Particle Properties advocates the use of the original form factor calculation by Leutwyler and Roos~\cite{Leutwyler:1984je}: $f_+(0)= 0.961 \pm 0.008$. Taking this value into account, the value of $V_{us}$ is found to be:

\begin{equation}
| V_{us} | = 0.257 \pm 0.0021 
\end{equation} 

\begin{figure}[h]
\centering
\includegraphics[width=80mm]{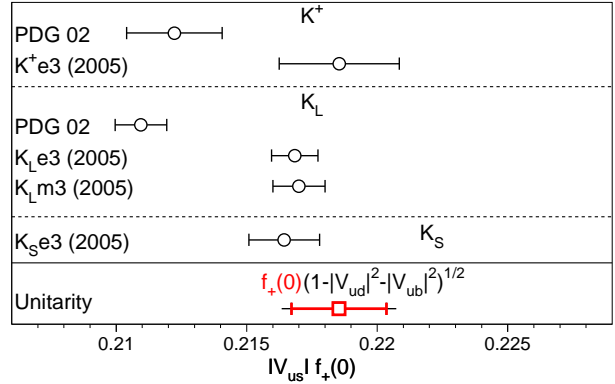}
\caption{Comparison of the determinations of $|V_{us}|f_+(0)$.} \label{bluchermarciano}
\end{figure}

A pictorial representation of the state-of-the art results is shown in Fig.~\ref{bluchermarciano}~\cite{pdgupdate}.
 In conclusion, over the past few years, the $V_{us}$ revolution which started with a re-analysis of hyperon decays~\cite{Cabibbo:2003ea} and the measurement of the semileptonic decay of the charged kaon by BNL-E865~\cite{Sher:2003fb} has led to an overall agreement between the results from neutral and charge semileptonic kaon decays, leptonic kaon decays, hyperon and tau decays! It this the end of the story or there will be more surprises? The onus is on the theory side to provide reliable form factor calculations.

\section{Search for Direct CP violation in Charged Kaon Decays}
\label{na48/2}
The CERN SPS has remained the only source of high energy secondary hadron beams. After the completion of the $\epsilon^\prime / \epsilon$ program and of the follow-up experiment (NA48/1) devoted to the study of rare $K^0_{S}$ decays, the NA48 Collaboration has collected data with two simultaneous charged beams of opposite polarity with the main objective to study direct CP violation in charged kaon decays into three pions. The experiment, known as NA48/2, has accumulated about 4 billion charged kaon three pion decays. $K^+$ and $K^-$ decays were collected simultaneously to avoid systematic errors related to the change of detector conditions as a function of time. The main motivation of NA48/2 was to compare the density of points in the Dalitz plot for $K^+$ and $K^-$ has a function of the Lorentz-invariant quantity $u = (s_3 - s_0)/m^2_{\pi}$. In the kaon rest frame, the two Lorentz-invariant variables $u$ and $v$ can be written as:
\begin{equation}
\begin{array}{c} 
u = 2 m_K \cdot \frac{(m_{K/3} - E_{odd})}{m^2_\pi} \\
v = 2 m_K \cdot \frac{(E_1-E_2)}{m^2_\pi},
\end{array}
\end{equation}
where $E_1$ and $E_2$ are the kinetic energies of the pions with charge equal to that of the parent kaon and $E_{odd}$ is the kinetic energy of the pion of opposite charge. 
The density of the Dalitz plot can be written as:
\begin{equation}
|M(u,v)|^2 \simeq 1 + gu + hu^2 + k v^2
\end{equation} 
A slope asymmetry
\begin{equation}
A_g = \frac{g_+-g_-}{g_++g_-}\simeq \Delta g /(2g)
\end{equation}
different from zero signals direct CP violation in charged kaon decays. In the equation above $g$ is the average slope and $\Delta g$ is the slope difference. The prediction of $A_g$ in the Standard Model are safely smaller than the achievable precision and the experiment, which has an order-of-magnitude better sensitivity than the previous ones, can  look for new CP violating effects, such as CP violation induced by SUSY chromomagnetic penguins~\cite{D'Ambrosio:1999jh}. 

\begin{figure*}[t]
\centering
\includegraphics[width=135mm]{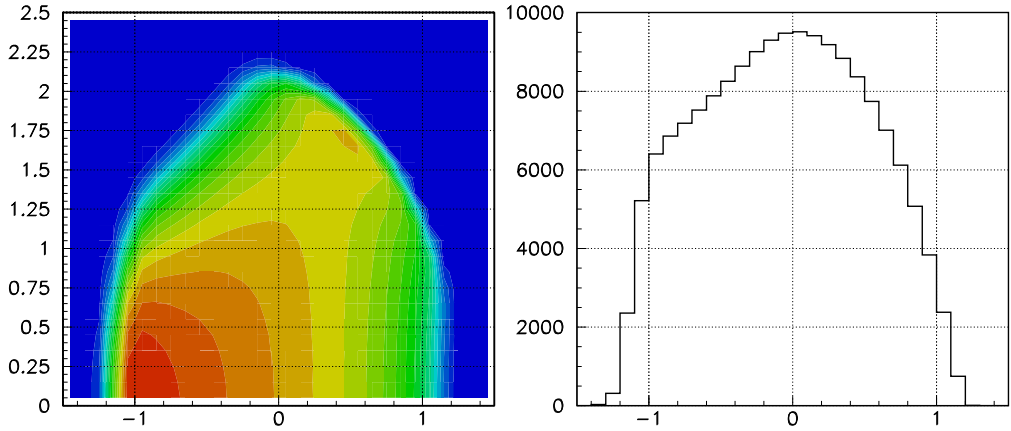}
\caption{Left: normalized distribution of the selected events in the kinematical variables $u$ and $|v|$; right: $u$ distribution of the selected events.} \label{uv}
\end{figure*}

To give an idea of the unprecedented quality of the charged kaon data analyzed by NA48/2, the Dalitz plot of the events collected during the year 2003 is shown in Fig.~\ref{uv}.

\begin{figure}[ht]
\centering
\includegraphics[width=80mm]{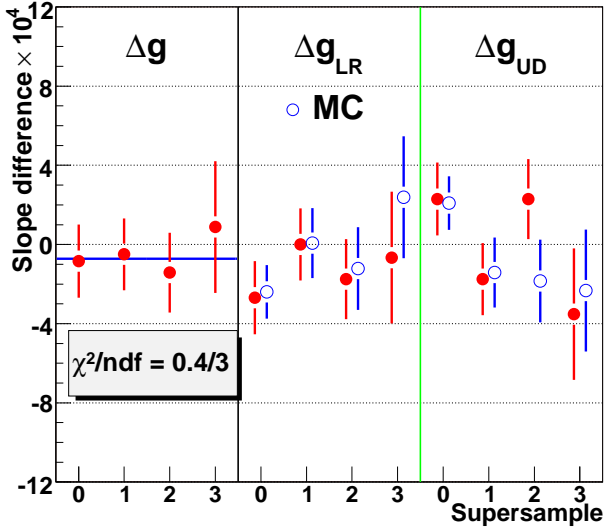}
\caption{Measurement of the slope difference $\Delta g$ for the four data subsamples and related control quantities.} \label{deltagmeas}
\end{figure}

 The measurement method is based on comparing the $u$ spectra of $K^+$ and $K^-$ decays $N^+(u)$ and $N^-(u)$. Given the actual values of the slope parameters $g$, $h$, and $k$ and the precision of the measurement, the ratio of $u$ spectra is in good approximation proportional to: 
\begin{equation} 
R(u) = \frac{N^+(u)}{N^-(u)} \simeq 1 + \Delta g \cdot u 
\end{equation} 
from which $A_g$ can be derived. 
To address possible beam-line asymmetries, the polarity of the beam achromat that setup was regularly exchanged. To compensate for possible left-right detector asymmetries, the polarity of the dipole magnet used to measure the pion momenta was frequently changed. Details about the analysis can be found in the publication~\cite{Batley:2006mu}. The Result of the $\Delta g$ measurement for the four subsample of data collected by NA48/2 in 2003 in shown in Fig.~\ref{deltagmeas}.

In conclusion, the difference in linear slope parameter of the Dalitz plot for 3$\pi^\pm$ decays of $K^+$ and $K^-$, measured with data collected in 2003 is found to be: 
\begin{equation}
\Delta g = (-0.7 \pm 0.9_{\rm stat} \pm 0.6_{\rm trig} \pm 0.6_{\rm syst}) \times 10^{-4}.
\end{equation}
Using the Dalitz plot slope $g = -0.2154 \pm 0.0035$~\cite{Eidelman:2004wy}, the measurement of the CP violating asymmetry is found to be compatible with zero and reads:
\begin{equation}
\begin{array}{c}
A_g = (1.7 \pm 2.1_{\rm stat} \pm 1.4_{\rm trig} \pm 1.4_{\rm syst}) \times 10^{-4} \\
= (1.7 \pm 2.9 ) \times 10^{-4}.  
\end{array}
\end{equation}

 The CP asymmetry can also be studied in the decays $K^\pm \to \pi^\pm \pi^0 \pi^0$. In this case the smaller branching ratio and acceptance for the final state is compensated by the larger Dalitz plot slope. The preliminary NA48/2 result based on data collected in 2003 yields: 
 
\begin{equation}
\Delta g_0 = (2.3 \pm 2.8_{\rm stat} \pm 1.3_{\rm trig} \pm 1.0_{\rm syst} \pm 0.3_{\rm ext}) \times 10^{-4}.
\end{equation}
Using the Dalitz plot slope $g_0 = 0.638$ the CP violating asymmetry reads:
\begin{equation}
\begin{array}{c}
A^0_g = (1.8 \pm 2.2_{\rm stat} \pm 1.0_{\rm trig} \pm 0.8_{\rm syst} \pm 0.2_{\rm ext}) \times 10^{-4} \\
= (1.8 \pm 2.6 ) \times 10^{-4}  
\end{array}
\end{equation}

These measurement are in agreement with the Standard Model. More data is available and the quest to seek CP violation beyond the Standard Model in charged kaon decays continues. New results should to be published soon.

\section{ Determination of the $\mathbf{\pi\pi}$ Scattering Length}
\label{cusp}

\begin{figure}[ht]
\centering
\includegraphics[width=120mm]{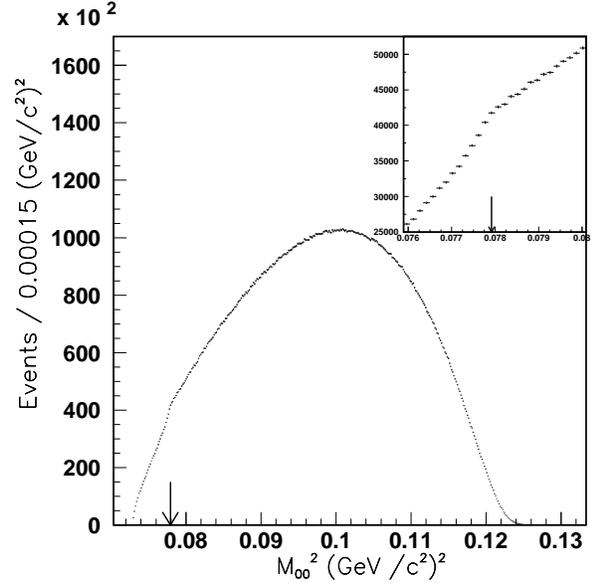}
\caption{Distribution of the square of the $\pi^0 \pi^0$ invariant mass. The insert is an enlargement of a narrow region centered at the $\pi^+ \pi^-$ threshold.} \label{figcusp}
\end{figure}

A very important --and totally unexpected! -- by-product of NA48/2 is the measurement of the $\pi \pi $ scattering length extracted from the $\pi^0 \pi^0$ invariant mass distribution of $K^\pm \to \pi^\pm \pi^0 \pi^0$ decays. The measurement is made possible by the excellent energy and position resolution of the NA48 Liquid Krypton calorimeter (LKr) and by the very large accumulated statistics ($\simeq 2.3 \times 10^7$ decays). The $\pi^0 \pi^0$ invariant mass distribution ($M^2_{00}$) is shown in Fig.~\ref{figcusp}. 

The cusp-like effect at $M^2_{00} = (2m_{\pi^+})^2=0.07792~ ({\rm GeV}/{\it c}^2)^2$ was never observed before. It is interpreted as $K^\pm \to \pi^\pm \pi^+\pi^-$ threshold effect contributing to the 
 $K^\pm \to \pi^\pm \pi^0\pi^0$ amplitude through the charge exchange reaction $\pi^+\pi^- \to \pi^0 \pi^0$. In the rescattering model of Cabibbo~\cite{Cabibbo:2004gq}, the $K^\pm \to \pi^\pm \pi^0\pi^0$ amplitude is described as the sum of two terms:
 \begin{equation}
 {\cal M}(K^\pm \to \pi^\pm \pi^0\pi^0) = {\cal M}_0 + {\cal M}_1, 
 \end{equation}
 where ${\cal M}_0$ is the unperturbed amplitude and ${\cal M}_1$ is the contribution from $K^\pm \to \pi^\pm \pi^+\pi^-$ decay through the $\pi^+ \pi^- \to \pi^0\pi^0$ charge exchange reaction given by:
 \begin{equation}
 {\cal M}_1 = -2a_xm_{\pi^+}{\cal M_+}\sqrt{1-\left(\frac{M_{00}}{2m_{\pi^+}}\right)^2}.   
 \end{equation}
 
 In the formula above, $a_x$ is the S-wave $\pi^+\pi^-$ charge exchange scattering length (threshold amplitude) and ${\cal M}_+$ is the known $K^\pm \to \pi^\pm \pi^0\pi^0$ decay amplitude at $M_{00}=2m_{\pi^+}$. ${\cal M}_1$ changes from real to imaginary at $M_{00}=2m_{\pi^+}$ and interferes destructively with ${\cal M}_0$ in the region $M_{00} < 2m_{\pi^+}$ while it adds quadratically above it. In the limit of isospin symmetry $a_x = (a_0-a_2)/3$, where $a_0$ and $a_2$ are the S-wave $\pi\pi$ scattering lengths in the $I=0$ and $I=2$ states, respectively. A more complete formulation of the 
model~\cite{Cabibbo:2005ez} takes into account all rescattering processes at the one- and two-loop level. Using the formalism of this improved description, the NA48/2 data yield~\cite{Batley:2005ax}:
\begin{equation}
(a_0-a_2)m_{\pi^+} = 0.268 \pm 0.010_{\rm stat.} \pm 0.004_{\rm syst.} \pm 0.013_{\rm ext.}. 
\end{equation} 
The additional external error of $\pm 0.013$ has been assigned to cover the branching ratio uncertainty and the theoretical error. The result is in good agreement with the latest theoretical prediction based on the framework of chiral perturbation theory~\cite{Colangelo:2000jc}: 
\begin{equation}
(a_0-a_2)m_{\pi^+} = 0.265 \pm 0.004. 
\end{equation}

\section{ Leptonic Kaon Decay and New Physics}
\label{susy}
The large kaon decay exposure accumulated by NA48/2 has revamped the interest to look with better precision to the two-body leptonic kaon decays which are suppressed by helicity conservation in the the Standard Model but generally unsuppressed in extensions of it. An example is given by the ratio $R_K$, defined as follows: 
\begin{equation}
R_K = \frac{\Gamma(K \to e \nu(\gamma))}{\Gamma(K \to \mu \nu (\gamma))}.
\end{equation}  
   A new measurement of $R_K$ based on data collected by NA48/2 during the 2003 SPS proton run was presented at the HEP2005 Europhysics conference in Lisbon~\cite{Luca}: 
\begin{equation}
R_K =  ( 2.416 \pm 0.043_{\rm stat.} \pm 0.024_{\rm syst.}) \times 10^{-5}. 
\end{equation} 
This value is to be compared with the theoretical prediction $R_K=(2.472 \pm 0.001) \times 10^{-5}$ and the present world average~\cite{Eidelman:2004wy}: $2.45 \pm 0.11 \times 10^{-5}$. The slight discrepancy between the measurement and the prediction has triggered theoretical interest. In a recent article~\cite{Masiero:2005wr}, the role of low energy Supersymmetric (SUSY) extensions of the Standard Model to $R_K$ was examined. In general the ratio can be written as:
\begin{equation}
R_K=\frac{\Gamma^{K \to e \nu_e}_{SM}}{\Gamma^{K \to \mu \nu_\mu}_{SM}}\left(1+\Delta r^{e - \mu}_{NP}\right).
\end{equation}
The authors point out that for some possible regions of the minimal supersymmetric Standard Model (MSSM) is possible to obtain $\Delta r^{e - \mu}_{NP}$ of the order of ${\cal O} (1 \%)$ and that such large contributions would arise from lepton flavor violation (LFV) effects. The quantity which accounts for the deviation from the $\mu - e$ universality is:
\begin{equation}
R^{LFV}_K = \frac{\sum_i \Gamma(K \to e \nu_i)}{\sum_i \Gamma(K \to \mu \nu_i)},
\end{equation}    
with the sum extended to all neutrino flavors because experimentally one determines only the flavor of the charged lepton. The dominant contributions to $R^{LFV}_K$ arise from the charged Higgs exchange and it is proportional to $\tan \beta^6$, to the fourth power of the charged Higgs mass and the square of the flavor changing parameter. 

\section{CKM Matrix and Rare Kaon Decays}
\label{ckmmat}
In the first section of this article we have described the progress on the determination of the magnitude of the CKM parameter $|V_{us}|$ from {\it frequent} tree kaon decays. We now address those rare kaon decays which allow us to learn short distance information and that, if the Standard Model is assumed, determine the magnitude and the phase of another CKM parameter, $V_{td}$. We will limit the discussion to $K \to \pi \nu \nu$ transitions because the theoretical predictions within the Standard Model are very precise.  The CKM matrix~\cite{Cabibbo:1963yz,Kobayashi:1973fv} is responsible for the quark mixing patterns and can be parameterized in a way to emphasize its hierarchical patterns in powers of the Cabibbo angle $\lambda = \sin \theta_C$. Following Wolfenstein\footnote{In the analysis one will actually use a more exact parameterization and the modified Wolfenstein parameters $\bar{\rho} = \rho(1-\lambda^2/2)$ and $\bar{\eta} = \eta(1-\lambda^2/2)$.} , it has become common to display the matrix as follows~\cite{Wolfenstein:1983yz,Buras:1994ec}: 

\begin{itemize}
\item
$V_{us} \simeq \lambda$

\item 
$V_{cb} \simeq \lambda^2 A$

\item 
$V_{ub} \simeq \lambda^3 A (\rho - i \eta)$ 

\item 
$V_{td} \simeq \lambda^3 A (1 - \rho - i \eta)$, 
\end{itemize}

and:

\begin{equation}\label{Wolfenstein}
V_{\rm CKM} = 
\pmatrix{ 1-\frac12\lambda^2 & \lambda & A\lambda^3(\rho-i\eta) \cr
  -\lambda & 1-\frac12\lambda^2 & A\lambda^2 \cr
  A\lambda^3(1-\rho-i\eta) & -A\lambda^2 & 1} .
\end{equation}

The unitarity of the CKM matrix can be expressed displaying the relations between the matrix elements by means of triangles in the $ \rho - \eta$ plane. It has become common to use the triangle whose sides are of comparable magnitude: 

\begin{equation} 
V_{ud}V_{ub}^* + V_{cd}V_{cb}^* + V_{td}V_{tb}^* =0 
\end{equation} 

The past decade has witnessed decisive tests of the pattern of quark mixing and and CP violation. Nonetheless the flavor sector still remains one of the least tested aspect of the Standard Model. The overall strategy to make further progress can be outlined as follows:

\begin{enumerate} 

\item

Determine the magnitude of the CKM parameters from semileptonic meson decays or from oscillations. Examples of this class of measurements are: 

\begin{itemize} 

\item{$|V_{us}|$}

$K^0_{L} \to \pi^\pm e^\mp \nu$ 

\item{$|V_{cb}|$}

$ B \to X_c \ell \nu$

\item{$|V_{ub}|$}

$ B \to X_u \ell \nu$

\item{$|V_{td}|$}

In the ratio of $B_{d,s}^0 - {\bar B_{d,s}^0}$ oscillations, the theoretical quantities needed from lattice QCD ($f_B$ and $\hat{B}$) enter only as ratios, thus the theoretical control is good enough to determine $V_{td}/V_{ts}$. 

\end{itemize}

\item 

Determine the phases of the CKM complex elements those CP-asymmetries least affected from hadronic uncertainties and insensitive to new physics.   

\begin{itemize} 

\item{$\arg{V_{td}} \equiv  \exp \left(-i \beta \right)$}

The cleanest determination of $\sin 2\beta$ is obtained the mixing-induced, time-dependent CP violating asymmetry in $B^0 ({\bar B^0}) \to J/\psi K^0_{S}$.

\item{ $\arg{V_{ub}} \equiv \exp \left( i \gamma \right)$} 

A tree level determination of $\gamma$ requires, for instance, the study of $B \to D K$ and $B_s \to D_s K$ decays. 
\end{itemize}

\item

Look for New Physics by studying highly suppressed Flavor Changing Neutral Current (FCNC) processes which proceed through loop diagrams. Decisive tests can be done if the Standard Model predictions are precise, that is immune from hadronic uncertainties, and the loop contributions are short distance dominated. These observables can be affected by contributions from new degrees of freedom in the loops and a deviation from the Standard Model predictions would signal New Physics. The best examples are given by rare kaon decays, $B$ leptonic decays and Forward-Backward asymmetries in $B$ decays:

\begin{itemize} 

\item 

$K^0 \to \pi^0 \nu \bar{\nu}$

\item 

$K^+ \to \pi^+ \nu \bar{\nu}$

\item 
$K^0_{L} \to \pi^0 \ell^+ \ell^-$

\item 

$B_{d,s} \to \mu^+ \mu^-$

\item

$A_{FB} ( B \to X_s \ell^+ \ell^-)$

\end{itemize}

\end{enumerate}

\section{ $\mathbf{K^0_{L} \to \pi^0 \nu \bar{\nu}}$} 
\label{E391a}

While the theoretical error of prominent FCNC processes, such us $B \to X_s \gamma$ and $B \to X_s \mu^+ \mu^-$, amount to $\pm 10 \%$ or larger, the irreducible theoretical uncertainty on the $BR(K^0_{L} \to \pi^0 \nu \bar{\nu})$ amounts to only 1-2\%~\cite{Bryman:2005xp}. The Standard Model prediction can be expressed as follows~\cite{pdgupdate}:
\begin{equation}
\begin{array}{c}
{\cal BR}(K^0_{L} \to \pi^0 \nu \bar{\nu}) =  \\
 7.6\times 10^{-5} \cdot |V_{cb}|^4\cdot \bar{\eta}^2 \simeq \\
 (3.0 \pm 0.6) \times 10^{-11} 
\end{array}
\end{equation}  
The error is completely dominated by $|V_{cb}|$ and by the value of $\bar{\eta}$ implied from the fits to the unitarity triangles. The error is therefore parametric in nature and not due to uncertainties related to theoretical errors or hadronic matrix elements. This miraculous situation derives from the fact that these decays are mediated by a single operator and the corresponding hadronic matrix element can be experimentally measured from the semileptonic tree decays. This situation is unique in the field of meson physics. This is true for both  $K^0_{L} \to \pi^0 \nu \bar{\nu}$ and $K^+ \to \pi^+ \nu \bar{\nu}$.  The experimental information on the $K^0_{L} \to \pi^0 \nu \bar{\nu}$ has recently been improved by the first dedicated attempt, experiment E391a at the KEK PS~\cite{Lim:2004sa}. 
Before we describe the experimental result obtained by E391a, we need to spend a few words on the experimental difficulty to measure such a theoretically clean but elusive decay. Obviously the neutrino-antineutrino pair cannot be measured. Conceptually one would like to know as much as possible about the incoming neutral kaon: for instance making the neutral kaons at a $\phi$ factory where particle have a well defined momentum would simplify the analysis. In addition, by reconstructing the direction accompanying short-lived neutral kaon, the line-of-flight of the $K^0_{L}$ is known and the initial state is fully constrained. Unfortunately, given the small cross-section to make $\phi$ mesons in $e^+ e^-$ annihilations, and the limited luminosity achievable at low energy colliders, there is just no way to envisage the study of very rare kaon decays at a $\phi$ factory. 
The only viable method to produce a large enough amount of $K^0_{L}$ is to make the kaons using a proton beam hitting a fixed target. As far as the kaon flux is concerned, the energy of the proton beam is not critical as long as it is high enough to make kaons with good efficiency (that is as long as the energy of the primary protons is well above the threshold to make kaons!). It is possible to determine the momentum of the neutral kaon by Time of Flight (TOF) but only if its momentum is kept to be less than 1 GeV/{\it c}. This was the method proposed by the KOPIO experiment~\cite{KOPIO}, which unfortunately will not be built. The TOF technique is very expensive in the use of protons because in order to make low energy kaons, the neutral beam needs to be taken at a large production angle. This implies that to collect a sufficient flux of neutral kaons, a very wide beam acceptance needs to be considered. In turn, the useful acceptance of practical detectors tend to be very small. 
  Currently, the only method being pursued to study $K^0_{L} \to \pi^0 \nu \bar{\nu}$ is the so called {\it pencil beam} technique. Pioneered by KTeV at Fermilab, it consist in making neutral kaons --together with many many neutrons and photons unfortunately!-- taking the neutral beam at a small production angle and to keep the acceptance of the beam small in order to be able to use the beam direction as kinematical constrain in the form of a minimum transverse momentum cut. This is useful, for instance, to reject $\Lambda \to n \pi^0$ decays where the neutron escapes detection and the $\pi^0$ could mimic the final state under study. 
  If the kaon momentum is unknown, the only handle available is to fully measure the $\pi^0$ in the final state and hermetically veto any other particle.  
  This method was proposed by KAMI~\cite{Alexopoulos:2001gz} at Fermilab and not considered viable by a technical review committee. Nevertheless, the same method, in a step-by-step approach is being applied by the  KEK E391a experiment at the 12 GeV PS and proposed for further studies at the J-PARC hadron facility. 
  Three E391a runs have taken place. Results from a fraction of the first one where presented by K.~Sakashita at the KAON2005 Conference~\cite{kaon2005}:
\begin{equation}  
\begin{array}{c}
{\cal BR}(K^0_{L} \to \pi^0 \nu \bar{\nu}) < 2.86 \times 10^{-7}\\ ~~90\% (C.L.).
\end{array}
\end{equation} 
  
  This result is far from the interesting region which the experiment aims to test. With the data collected so far, there is hope to reach the Grossman-Nir model-independent upper bound~
\cite{Grossman:1997sk}:
\begin{equation}
\begin{array}{c}
{\cal BR}(K^0_{L}  \to \pi^0 \nu \bar{\nu})  \simeq \frac{\tau_{L}}{\tau_{K^+}} {\cal BR}(K^+ \to \pi^+ \nu \bar{\nu})\\ \simeq 1.4 \times 10^{-9}~~(90\% C.L.)    
\end{array}
\end{equation} 
A Letter of Intent to continue studies at J-PARC has been submitted~\cite{LOI05}. 
  
 \section{$\mathbf{K^+ \to \pi^+ \nu \bar{\nu}}$}
 \label{KPLUS}
 
 The situation concerning the charged kaon mode is radically different because there are already experimental results with the sensitivity of the Standard Model. It should also be clarified that the physics of the $K^+ \to \pi^+ \nu \bar{\nu}$ mode differs from the corresponding neutral kaon:
 \begin{itemize}
 \item
 The decay is not dominated by CP violating effects. 
 \item
 Owing to the charm contribution which entails uncertainties related to the quark masses and scale uncertainties, the prediction, although very precise,  cannot match the accuracy of  the $K^0_{L}$ mode.  
 \end{itemize} 
  The two modes are complementary and their simultaneous measurement would allow us to determine $\sin 2\beta$ from loop processes in a completely independent way, providing a decisive test of the Standard Model. The Standard Model prediction can be parameterized as follows~\cite{pdgupdate}: 
  
  \begin{equation}
  \begin{array}{c}
  {\cal BR}(K^+ \to \pi^+ \nu \bar{\nu}) \simeq  
 1.6\times 10^{-5} \cdot \\ |V_{cb}|^4\cdot \left( \sigma \bar{\eta}^2 + (\rho_c-\bar{\rho})^2\right) \simeq 
 (8.0 \pm 1.1) \times 10^{-11}, \\
 \sigma = 1 / (1- \lambda^2/2)^2, \\
 \rho_c \simeq 1.4.
  \end{array}
  \end{equation} 
 
  The recent NNLO calculation~\cite{Buras:2005gr} has reduced the purely theoretical error to an extend that will not be matched by experiments for some time. In fact on the experimental side, thanks to the endeavor at BNL, where the decay was studied with stopped kaons by E787 and its upgraded version, E949, progress has been steady over the years, leading to the publication of three candidates and to a first determination of the branching ratio~\cite{Anisimovsky:2004hr}: 

\begin{equation}
\begin{array}{c}
{\cal BR}(K^+ \to \pi^+ \nu \bar{\nu}) = \\
(1.47^{+1.30}_{-0.89}) \times 10^{-10}.
\end{array}
\end{equation}
The mean value is about two times larger than the Standard Model prediction but, given the large errors, completely in agreement with it. Unfortunately E949 was terminated prematurely and we are not learning as fast as possible about this interesting decay. Plans to pursue the stopped kaon technique exist in Japan in the form of a Letter of Intent~\cite{LOI04}. In the long run,  
to be useful, further studies shall aim to accumulate order-of-magnitude larger event samples in order to precisely measure the strength of the $s \to d$ transitions. This was the purpose of Fermilab proposal called {\it Charged Kaons at the Main Injector} or CKM for short~\cite{CKMexp}. The proposal intended to study decays in flights from a high momentum (21 GeV/{\it c}) separated kaon beam. To keep the background small, redundant measurements of the momentum of the incoming kaon and outgoing pion were to be done employing magnetic and velocity (RICH) spectrometers, with photon vetoes surrounding the decay tank to veto $\pi^0$ from $K^+ \to \pi^+ \pi^0$ decays and photons from radiative decays. The experiment was approved five years ago but never funded and eventually terminated. 

\begin{figure}[h]
\centering
\includegraphics[width=80mm]{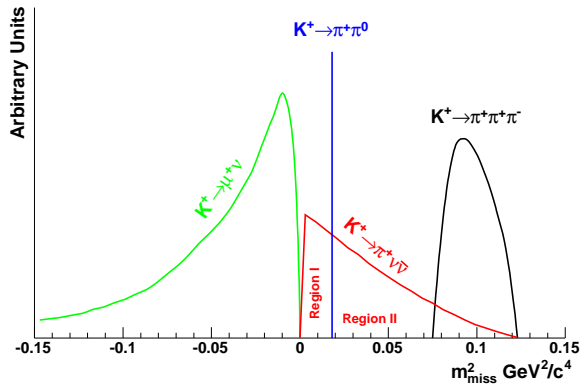}
\caption{The missing mass constraint to separate $K^+ \to \pi^ \nu \bar{\nu}$ from frequent kaon decays.} \label{kinem}
\end{figure}

To be effective in reducing background coming from $K^+ \to \pi^+ \pi^0$ where the $\pi^0$ is lost and the angle between the $K^+$ and the $\pi^+$ is mistakenly reconstructed, the use of even higher kaon beam momentum is preferable because the capability to detect photons with calorimetric detectors improves as a function of the energy. This is the consideration at the base of proposal P-326 for the CERN SPS. The CERN SPS delivers protons of 400 GeV/{\it c}. To optimise the number of useful decays, the secondary beam is chosen to have 75 GeV/{\it c}. Requiring the $\pi^+$ to have less than 35 GeV of energy, leaves at least 40 GeV of energy associated to the $\pi^0$. Such large amount of electromagnetic energy cannot be missed in the calorimeter! The importance to suppress the two body decay $K^+ \to \pi^+ \pi^0$ can be appreciated from Fig.~\ref{kinem} where it is shown how this decay forces one to split the signal acceptance in two missing mass regions.

The drawback of this design is that the momentum of the secondary beam is too high to separate kaons from protons and pions by means of radio frequency cavities. Although only about 6\% of particles are kaons and only $\simeq$ 10\% of the kaons decay in the useful decay volume, all particles (800 MHz) have to be tracked and precisely timed. This is the main challenge of the proposal. 

In addition to the availability of the SPS, which will definitely operate in the future because it is needed to inject protons in the LHC, the proposal builds on the experience and infrastructure of NA48. In particular, the experimental hall, the target, the beam line, the decay tank, the spectrometer dipole magnet and the Liquid Krypton calorimeter will be used in the new experiment. 
The main characteristics of the beam are: 

\begin{itemize} 
\item
rotons on target: $3 \times 10^{12}$ per cycle;
\item
beam rate: 800 MHz; 
\item
kaon decays: $\simeq$ 5 MHz decays of $K^+$ decays in the fiducial volume; 
\item 
duty cycle 4.8 s / 16.8 s = 0.29; 
\item
useful kaon flux/year: 5$\times 10^{12}$.
\end{itemize}

The main elements of the proposed detector are:
\begin{itemize} 
\item
A differential Cherenkov counter (CEDAR) for positive kaon identification
\item 
A beam tracker (GIGATRACKER) formed by two stations of silicon micro-pixels with excellent timing capability and a micromegas-based TPC as already employed in NA48/2 (KABES) to track the upstream particles. The gaseous detector in the last station is meant to minimize the multiple scattering and the beam interactions before the particles enter the decay region.
\item 
Photon anti counters (ANTI) to give hermetic coverage up to 50 mrad for photons originating from kaon decays in the fiducial volume. 
\item
A system of six straw tracker stations (STRAW) operated in vacuum to minimize multiple scattering. The momentum of the charged tracks is measured twice using two large dipoles. One dipole already exists from NA48. 
\item
The Liquid Krypton calorimeter of NA48 (LKr). This is one of the best photon detector ever built in HEP. Its future use as photon veto is not a demotion: the excellent position time and energy resolution are essential ingredients to actually measure the photon detection efficiency and make a reliable estimation of the backgrounds.  
\item
An 18 m long neon RICH for muon-pion separation. 
\item
The NA48 hodoscope for fast timing and triggering (CHOD). The precise timing of the charged pion is obtained combining information from the RICH and the CHOD.
\item
A magnetized muon detector and hadron calorimeter (MAMUD) made of iron plates and extruded scintillator. The magnetization is provided by two coils and it is meant to sweep the charged beam away from the photon veto that complements the photon coverage at the end of the hall.
\end{itemize}

 The proposal~\cite{Anelli:2005ju} was submitted in June 2005. It is know by the sequential number P-326 but also as NA48/3 because it builds on the expertise and infrastructure of NA48. After being reviewed by the CERN scientific committee (SPSC), the project was endorsed by CERN as R\&D. The possibility to collect 40 signal events per year (for a branching ratio of $10^{-10}$) is based on time estimates and efficiency figures taken from the decennial experience of NA48 at the SPS and therefore it is completely realistic. There is not lack of opportunity, for motivated groups, to join the Collaboration! 

\bigskip 
\begin{acknowledgments}
I thank the organizing committee for the excellent conference, rich of exchange, discussion and....new results! A special mention to D.~Bryman, who I wish to see soon at CERN to discuss the new rare decay experiment at the SPS. Last but not least, skiing in Whistler and Blackcomb Mountain will definitely take me back to British Columbia. 
\end{acknowledgments}

\bigskip 

\end{document}